# Calorimetric Investigation of NdBa$_2$Cu$_3$O$_x$ Single Crystals

U. Tutsch[1], P. Schweiss[2], R. Hauff[1], B. Obst[1], Th. Wolf[1], and H. Wühl[1,3]

*Forschungszentrum Karlsruhe, [1]ITP and [2]INFP, 76021 Karlsruhe
and [3]Universität Karlsruhe, IEKP, 76128 Karlsruhe, Germany*

*The specific heat of Nd(Y)Ba$_2$Cu$_3$O$_x$ single crystals without substitution of Nd for Ba has been measured for various oxygen contents. In comparison with YBa$_2$Cu$_3$O$_x$, the maximum attainable transition temperature of the crystals is about 4 K higher and shifted from x = 6.92 to x = 7.0, the highest possible oxygen content. In spite of the difference in the $T_c(x)$ dependences, the jump in the specific heat $\Delta C/T_c$ at $T_c$ in both systems increases continuously being maximum at x = 7.0. Bond valence sum calculations based on neutron diffraction data point to a retarded generation of holes with increasing oxygen content in the Nd system and a different hole distribution between the chains and the planes in comparison to YBa$_2$Cu$_3$O$_x$ accounting for the different behaviour of both systems.*
PACS numbers: 74.25.Bt, 74.62.-c, 74.72.-h

## 1. Introduction

The YBa$_2$Cu$_3$O$_x$ high–$T_c$ family in many respects exhibits a systematic behaviour in its dependence on the oxygen content x in the CuO chains. When substituting Rare Earth ions for Yttrium, the maximum transition temperature $T_{c,max}$ at optimum oxygen concentration $x_{opt}$ increases with increasing ion radius and $x_{opt}$ is shifted to higher oxygen content.[1] As function of the hole density $n_{pl}$ in the CuO$_2$ planes one would expect a universal behaviour.[2] To check this, the hole distribution over the structural components of the unit cell has been evaluated from the bond lengths obtained by neutron diffraction analysis for both systems. The transition temperature and the condensation energy of the superconducting pairs have been obtained by specific heat measurements of NdBa$_2$Cu$_3$O$_x$ (Nd-123) for dif-ferent oxygen contents. They are compared with earlier results for Y-123.[3] Differences in superconducting properties are attributed to the different hole distribution within the unit cell.

## 2. Experimental

Nd-123 single crystals have been grown from CuO–BaO fluxes by the slow cooling method in Y–stabilised $ZrO_2$ crucibles at a reduced oxygen partial pressure in order to suppress an occupation of the Ba site by Nd. For the oxidation the crystals were annealed under flowing oxygen or oxygen/argon mixtures. The oxygen content x has been adjusted using the $x(T_{anneal})$ isobars of Lindemer et al.[4] Oxygen contents close to x = 7 were obtained by annealing the crystals at an oxygen pressure of 158 bar resulting in $T_c$ = 95.5 K. All crystals were heavily twinned.

Neutron–diffraction data (see below) of crystals used for the calorimetric investigations proved that the Cu sublattice as well as the apical oxygen sites were fully occupied. Thus, the oxygen contents of two crystals could be determined with high accuracy as x = 6.99 and x = 6.90, in excellent agreement with the pre–set oxidation conditions. There was no indication of a Nd substitution for barium. EDX analysis certified the purity of the crystals except for an 8 at% Y occupation of the Nd site due to corrosion of the crucible. In the following this is indicated by the notation Nd(Y)-123. The high quality of the single crystals had previously been demonstrated by the low critical current densities of crystals at maximum O content.[5] In addition, some tiny high quality Y–free crystals grown in $BaZrO_3$ crucibles were investigated with respect to $T_c$.

The specific–heat measurements were performed by a continuous heating technique on samples of about 20–40 mg. The specific heat of an oxidized Nd/Ba substituted crystal with a $T_c$ below 50 K, which is below the transition temperatures of the investigated crystals, was used as the normal–state background. A mean–field type specific–heat jump $\Delta C/T_c$ at $T_c$ has been obtained, analysing the superconducting transition near $T_c$ as 3D–xy critical fluctuations which exactly holds for Y-123 near optimal doping.[6,7]

The hole distribution between the $CuO_2$ planes, the apical oxygens, and the $CuO_2$ chain has been evaluated from bond valence calculations. Bond lengths were determined from neutron–diffraction data. The measurements were taken at the four–circle diffractometer $5C_2$ at the Orphee reactor,[8] using neutrons of 0.83 Å wavelength. For the orthorhombic twinned crystals the scan range was adjusted individually in order to achieve a complete integration over the typical multi–peak structure. Internal R values of the averaging procedure are about 2% for all data sets and between 2–3% for the refinement. From the resulting bond lengths and the oxygen occupation the bond valence sums at the Cu sites have been calculated. The apparent oxidation states of the Cu ions were corrected for internal stresses in order to ensure that the charges summed up over the three Cu ions in the unit cell meet the charge demanded by the composition.

## 3. Results

Fig.1 shows the calculated hole densities at the CuO$_2$ planes, apical oxygens, and CuO chain as function of x. For Y-123 the results mirror the influence of the ortho-II phase in Y-123 leading to the well–known $T_c(x)$ dependence with a plateau around x = 6.7. They are in good agreement with results of site–specific x–ray absorption spectroscopy by Merz et al.[9] This gives us confidence in our method of calculating the bond valence sums. For the optimum hole density they obtained 0.2 in comparison with our value of 0.24 (holes per $CuO_2$ plane). These values exceed the commonly used $n_{opt} = 0.16$ (see Ref.2). For a relative comparison of the two systems this difference is, however, not decisive.

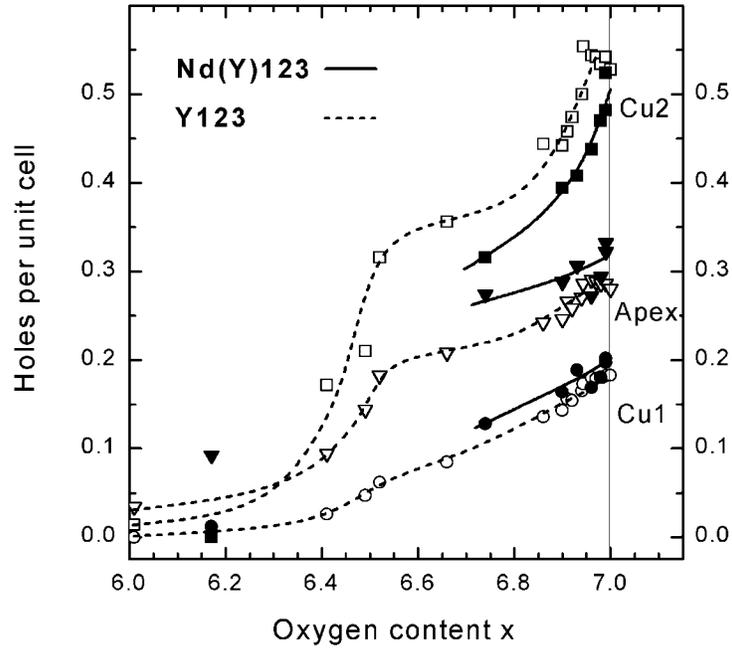

Fig. 1. Distribution of the holes over the CuO$_2$ planes, the apical oxygens, and the CuO chain of the unit cell of YBa$_2$Cu$_3$O$_x$ and Nd(Y)Ba$_2$Cu$_3$O$_x$ obtained from calculations of bond valence sums.

The calculations for Nd(Y)-123 show a different hole distribution within the unit cell in comparison to Y-123. More holes stay in the $CuO_3$ block formed by the CuO chain and the apical oxygens. The reduced hole density in the $CuO_2$ plane results in the shift of $x_{opt}$ from 6.92 (Y-123) to 7.0 for Nd(Y)-123 as shown in Fig.2. Consequently, $T_{c,max}$ of Y–free Nd-123 is further shifted to a fictitious x > 7.0. This retarded hole generation is in agreement with NQR results.[10] Unlike the behaviour of Y-123 where the oxygen ions in the chains preferentially order into long chain fragments, in Nd-123 the lattice is expanded by the bigger Nd ion and much shorter chains are formed. This has been attributed to a repulsive interaction between the oxygen ions leading to a more random occupation of the oxygen sites.[10] Thus, only at high oxygen contents when the chain fragments are bridged by oxygen ions, holes are introduced to the unit cell to a greater extent. This retarded hole generation in the Nd system is also expressed in the steep rise of the specific heat jump $\Delta C/T_c$ for x approaching 7.0.

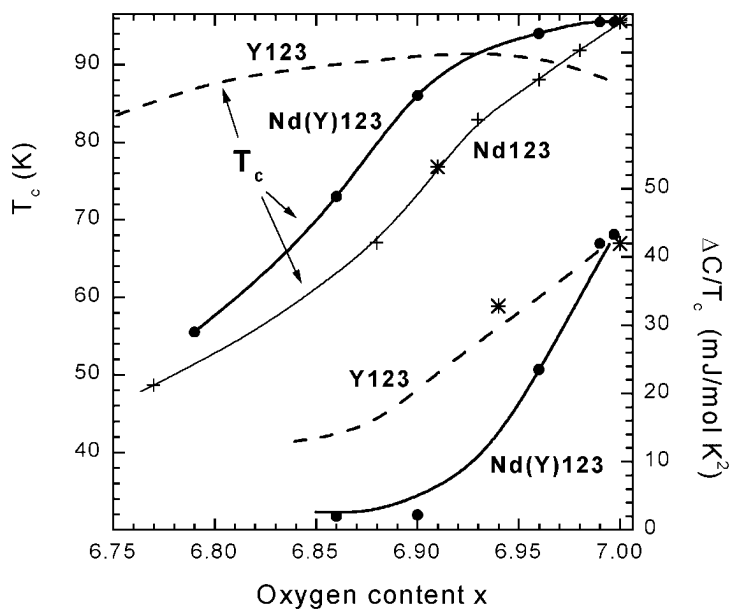

Fig. 2. Transition temperature and specific heat jump at $T_c$ for $YBa_2Cu_3O_x$ and $NdBa_2Cu_3O_x$. The $T_c$ values for Y–free Nd123 (+) are taken from Erb et al.[11] and two own values (*) are added. The jumps $\Delta C/T_c$ obtained from a 3D–xy analysis of the critical fluctuations are compared with earlier mean–field jumps[3] for Y123, treated in Gaussian approximation. For comparison, two of those specific heat transitions have been re–analysed in terms of 3D–xy critical fluctuations (*).

The enhanced condensation energy $\Delta C \cdot T_c$ indicates improved superconducting properties of NdBa$_2$Cu$_3$O$_{7.0}$. The high hole concentration in the CuO$_3$ blocks and a reduced inter–plane distance, due to the internal pressure of the large ion radius of the Nd ion, result in a stronger Josephson coupling of the planes across the CuO$_3$ blocks.[12] The system becomes more three-dimensional. The importance of the CuO$_3$ blocks for the appearance of superconductivity has recently been shown for Ca–doped Y-123 single crystals.[9] This situation might also be responsible for the high irreversibility field $B_{irr} = 13.4$ Tesla at 77 K for NdBa$_2$Cu$_3$O$_{7.0}$ compared to 10 Tesla for YBa$_2$Cu$_3$O$_{7.0}$.[5]

In conclusion, the maximum transition temperature of Nd(Y)Ba$_2$Cu$_3$O$_x$ appears at the expected hole concentration in the CuO$_2$ plane at x = 7.0. Its value and the superconducting strength can be explained by the hole distribution over the structural components.

## REFERENCES

1. M. Buchgeister et al., *Progress in High Temp. Supercond.*, ed. R. Nicolsky Vol.**25**, 511 (1990).
2. J.L. Tallon et al., *Phys. Rev. B* **51**, 12911 (1995).
3. V. Breit et al., *Phys. Rev. B* **52**, R15727 (1995).
4. T.B. Lindemer et al., *Physica C* **255**, 65 (1995)
5. Th. Wolf et al., *Phys. Rev. B* **56**, 6308 (1997).
6. V. Pasler et al., *Phys. Rev. Lett.* **81**, 1094 (1998).
7. A. Junod et al., *Physica C* **317-318**, 333 (1999).
8. Laboratoire Léon Brillouin, CE Saclay, Laboratoire Commun CEA–CNRS.
9. M. Merz et al., *Phys. Rev. Lett.* **80**, 5192 (1998).
10. H. Lütgemeier et al., *Physica C* **267**, 191 (1996).
11. A. Erb et al., *Applied Superconductivity, Proc. of EUCAS* 1997, 1109 (1997).
12. J.L. Tallon et al., *J. Low Temp. Phys.* **105**, 1379 (1996).